\documentclass[12pt,a4paper]{article}
\usepackage{jheppub}  

\usepackage{graphicx}
\usepackage{amsmath}
\usepackage{amssymb}
\usepackage{epsfig}
\usepackage{subcaption}
\usepackage{xcolor}
\usepackage{braket}
\usepackage{esint}
\usepackage{xcolor}

\newcommand{\dd}{\mathrm{d}}

\newcommand{\e}{\mathrm{e}}

\title{Higher rank Wilson loops in the $\mathcal{N}=2$ $SU(N)\times SU(N)$ conformal quiver}
\author{Benjo Fraser}
\affiliation{ 
 Department of Nuclear and Particle Physics\\
  Faculty of Physics\\
  University of Athens\\
  Athens 15784, Greece
}
\emailAdd{bjfraser@phys.uoa.gr}
\abstract{In this note we compute the expectation value of a circular BPS Wilson loop in the ``higher rank" totally symmetric and antisymmetric representations of $SU(N)$ in the $\hat{A}_1$ quiver $\mathcal{N}=2$ SCFT, using a matrix model. We discuss the connection with a recent conjecture stating that expectation values of observables in this sector are obtained from $\mathcal{N}=4$ SYM by a universal renormalization of the 't Hooft coupling. }
\begin{document}
\maketitle
\section{Introduction}
$\mathcal{N}=4$ super Yang-Mills theory (SYM) is the unique maximally supersymmetric theory in four dimensions. The only things we can `tune' in this theory are the gauge group $G$, and the gauge coupling constant $g_{YM}$. Arguably the next most special class of quantum field theories are the $\mathcal{N}=2$ superconformal field theories (SCFTs). By contrast with $\mathcal{N}=4$ SYM, these are numerous - significant progress has been made in their classification over last few years, including the discovery of the `class $\mathcal{S}$' theories \cite{Gaiotto:2009we}. It is an important research program to classify and understand them - see e.g. \cite{Tachikawa:2013kta} and references therein. The problem of obtaining gravity duals of such theories has been tackled in \cite{Grana:2001xn,Lin:2004nb,Gaiotto:2009gz,Gadde:2009dj,ReidEdwards:2010qs,Colgain:2011hb,Aharony:2012tz,Stefanski:2013osa}.

$\mathcal{N}=2$ SCFTs which have a Lagrangian description in terms of $SU(N)$ gauge groups have an ADE classification \cite{Kachru:1998ys,Lawrence:1998ja}, that is a one-to-one correspondence with affine simply-laced Lie algebras. Each theory takes the form of an $SU(N)$ quiver whose quiver diagram is the same as the Dynkin diagram for the corresponding algebra. By turning off gauge couplings for some of the nodes we can obtain theories corresponding to the finite ADE algebras. These theories are orbifolds of $\mathcal{N}=4$ SYM, and their gravity duals are therefore well known - they are orbifolds of $AdS_5\times S^5$ which leave invariant an $AdS_5\times S^1$ submanifold. 

For ADE theories with $K$ nodes, in the planar limit, following \cite{Gadde:2010zi,Pomoni:2011jj,Liendo:2011xb} it was conjectured in \cite{Pomoni:2013poa} that expectation values of operators in the so-called `$SU(2,1|2)$ sector', which in particular only involve fields in one vector multiplet of the quiver, were related to their $\mathcal{N}=4$ counterparts by a finite renormalization $g^2 \mapsto g_{\rm eff}^2 (g_1^2,\cdots g_K^2)$ of the coupling, where throughout for any gauge coupling we define e.g. $g^2\equiv g_{YM}^2 N/(4\pi)^2$. 

When formulated on $S^4$, $\mathcal{N}=2$ field theories are amenable to exact results using the localization technique, which was pioneered in this context in \cite{Pestun:2007rz}. For a certain class of supersymmetric observables, this reduces the computation of their expectation values to a matrix model. Although in general these models still cannot be solved for all values of the coupling, they are often soluble in the large-$N$ limit at either strong or weak 't Hooft coupling, and in the latter case provide a very efficient method for computing perturbative expansions - see important done in \cite{Russo:2012ay,Buchel:2013id,Russo:2013qaa,Russo:2013kea,Russo:2013sba}. Studying the matrix models for the $\mathcal{N}=4$ theory and the $\hat{A}_1$ quiver, it is possible to compute the conjectured universal coupling substitution.

In \cite{Mitev:2014yba}, the expectation value of a Wilson loop wrapped around the equator of $S^4$ in the first gauge group and in the fundamental representation was used for this purpose. It was found that
\begin{align}
\braket{W_{\Box}}_{\mathcal{N}=2} (g_1^2,g_2^2) = \braket{W_{\Box}}_{\mathcal{N}=4} (g_{\rm eff}^2(\Box) (g_1^2,g_2^2))
\end{align}
where $\braket{W_{\Box}}_{\mathcal{N}=4} (g^2)\, =\, I_1 (4\pi g)/(2\pi g)$ is the expectation value in the $\mathcal{N}=4$ theory. The effective coupling they found has the expansion has the expansion \cite{Mitev:2014yba}
\begin{align}
\label{eq:couplingsub}
g_{\rm eff}^2(\Box) (g_1^2&,g_2^2) = \nonumber\\
&g_1^2-12 g_1^4 \left(g_1^2-g_2^2\right) \zeta (3)+40 g_1^4 \left(g_1^2-g_2^2\right)\left(3 g_1^2+g_2^2\right) \zeta (5)+\,\cdots 
\end{align}

It is the proposal that the renormalization \eqref{eq:couplingsub} should be \textit{universal} for all operators in the $SU(2,1|2)$ sector, so that we could write $g^2_{\rm eff}(\Box)=g_{\rm eff}^2$. In this note we study a class of such operators: Wilson loops in representations of the gauge group other than the fundamental. In particular we consider the totally (anti-)symmetric rank-$k$ representation, where $k$ scales with $N$. For simplicity we concentrate on the $\hat{A}_1$ theory. `Higher rank' Wilson loops are important observables which can probe aspects of the theory that fundamental loops cannot. In $\mathcal{N}=4$ they have been studied extensively in the context of AdS/CFT \cite{Drukker:2005kx,Yamaguchi:2006te,Hartnoll:2006hr,Yamaguchi:2006tq,Gomis:2006sb,Okuyama:2006jc,Hartnoll:2006is,Gomis:2006im,Faraggi:2014tna}, while in the $N_f=2 N$ theory (which is the limit of the $\hat{A}_1$ quiver as one coupling goes to zero) they yield important information about any possible AdS dual \cite{Fraser:2011qa}. Very recently \cite{Chen-Lin:2015dfa}, they have been considered in $\mathcal{N}=2^*$ SYM, where they were used to probe phase transitions in the theory. 

%
%
%
%
%
%
%
\section{Higher rank Wilson loops}
In the $\hat{A}_1$ quiver theory, the supersymmetric Wilson loop operator in representation ${\cal R}$ is built out of the fields in one of the two vector multiplets:
\begin{align}
\label{wilsonloop}
W_R=\frac{1}{{\rm \dim}(\mathcal{R})}{\rm Tr}_{\cal R} \mathcal{P}\exp {\rm i}\int \dd x^{\mu}\left( A_{\mu}+ {\rm i}\, n_I\phi^I\, \frac{\dot{x}_{\mu}}{|\dot{x}|}\right) 
\end{align}
where $\phi^I$ ($I=1,2$) are the two real adjoint scalar fields in the vector multiplet, and $\mathbf{n}$ is a unit vector in $\mathbb{R}^2$. We can choose the first vector multiplet without loss of generality because the quiver theory has the $\mathbb{Z}_2$ symmetry $\{SU(N)_1\leftrightarrow SU(N)_2$, $g_1\leftrightarrow g_2\}$. The contour is around the equator of $S^4$. Note that we have normalized the operator using the dimension of the representation: ${\rm \dim}(S_k)=\tfrac{(N+k-1)!}{k!(N-1)!}$, ${\rm \dim}(A_k)=\tfrac{N!}{k!(N-k)!}$. The computation of the expectation value of \eqref{wilsonloop} in $\mathcal{N}=2$ theories on $S^4$ reduces to a matrix model for the constant modes of one of the adjoint scalar fields $a$, $b$ in each vector multiplet, which can be expressed as an integral over the eigenvalues $a_i,b_r$ ($i,r=1,\cdots N$):
\begin{align}
\label{matrixpath}
\langle W_R \rangle\,=\, &\frac{1}{{\rm dim}({\cal R})}\int \left(\prod_{i,r} \dd a_i\dd b_r\right)\,\e^{-\frac{N}{2}(\frac{1}{g_1^2}\sum_i a_i^2+\frac{1}{g_2^2}\sum_r b_r^2)}\nonumber\\
&\quad \prod_{i<j} (a_i-a_j)^2\, \prod_{r<s} (b_r-b_s)^2\,{\cal Z}_{\rm 1-loop}(a,b)\, |{\cal Z}_{\rm inst}(a,b)|^2\,{\rm Tr}_{\cal R}\e^{2\pi a}\nonumber\\
{\cal Z}_{\rm 1-loop}(a,b)\equiv &\,\frac{\prod_{i < j}H^2(a_i-a_j)\prod_{r < s}H^2(b_r-b_s)}{\prod_{i,r} H^2(a_i-b_r)}
\end{align}
where $H(x)\equiv \sum_{n=1}^{\infty}\left( 1+\frac{x^2}{n^2}\right) ^n \e^{-x^2/n} $. 
Passing to the large-$N$ limit, this integral is dominated by a continous saddle point distribution of eigenvalues for each of the scalars, which can be described by two eigenvalue densities $\rho_{1,2} (x)$. It can be shown \cite{Passerini:2011fe} that the instanton contribution $|{\cal Z}_{\rm inst}|^2$ is sub-leading in $N$, so we can safely ignore it in what follows. As is often the case, it is consistent to assume that $\rho_{1,2} (x)$ each have support on only a finite interval $[-\mu_{1,2},\mu_{1,2}]$, which we can take to be symmetric about the origin because of the $(a_i,b_r)\rightarrow (-a_i,-b_r)$ symmetry of the integral \eqref{matrixpath}.  The equations determining the distributions are then
\begin{align}
\label{evsaddles}
\fint_{-\mu_1}^{\mu_1} \dd y\, \rho_1 (y)\, \left( \frac{1}{x-y} - K(x-y)\right)+\fint_{-\mu_2}^{\mu_2} \dd y\, \rho_2 (y) K(x-y)\, &=\, \frac{1}{2 g_1^2}\, x\nonumber\\
\fint_{-\mu_2}^{\mu_2} \dd y\, \rho_2 (y)\, \left( \frac{1}{x-y} - K(x-y)\right)+\fint_{-\mu_1}^{\mu_1} \dd y\, \rho_1 (y) K(x-y)\, &=\, \frac{1}{2 g_2^2}\, x\\
\int_{-\mu_{1,2}}^{\mu_{1,2}} \rho_{1,2}\, (x)\, &=\, 1\nonumber
\end{align}
which explicitly display the $\mathbb{Z}_2$ symmetry $1 \leftrightarrow 2$ of the quiver. The third equation is the normalization condition for the densities. Here $K(x)\equiv -H'(x)/H(x)$, which has the expansion about $x=0$
\begin{align}
\label{kexp}
K (x)\, =\, -2\sum_{n=1}^{\infty} (-1)^n \zeta (2n+1) x^{2n+1}
\end{align}
which has the interpretation as the generating function for the class of `fan' superspace Feynman diagrams \cite{Broadhurst:1985vq,Mitev:2014yba}. In fact a fan diagram with $n$ faces and momentum $p$ running in the loop, together with its combinatorial factor, gives
\begin{align}
\label{fandiagrams}
\text{fan}_n (p)\, =\, 2\binom{2n-1}{n} \zeta (2n-1)\frac{1}{p^2}\, =\, 6\zeta (3)\frac{1}{p^2}\, ,\, 20\zeta (5)\frac{1}{p^2}\, \cdots\quad .
\end{align}
We will take two different field theory limits: one in which both 't Hooft couplings go to zero together, and another in which they become large together. These are two interesting limits since they correspond to a regime where the string orbifold is a useful description, but are not the only limits we could take - see the discussion section. In order to reach these regimes we use the parametrization $g_{1,2}\equiv\lambda\,\kappa_{1,2}$, with $\kappa_1^2+\kappa_2^2=1$, so that $\lambda^2=g_1^2+g_2^2$ can be thought of as the average coupling. Then, keeping both $\kappa_{1,2}$ of order one, we take $\lambda\rightarrow 0$ (`weak coupling') or $\lambda\rightarrow \infty$ (`strong coupling'). At weak coupling the equations \eqref{evsaddles} are solved by densities $\rho_{1,2} (x)$ and endpoints $\mu_{1,2}$ that take the form of Wigner semi-circle distributions with polynomial corrections:
\begin{align}
\rho_{1,2} (x)\, &=\, \frac{1}{\sqrt{\lambda}}\left(\sum_{\alpha,n=0}^{\infty}\rho_{1,2}^{(\alpha,n)}x^{\alpha}\lambda^n\right)\sqrt{\mu_{1,2}^{\,2}-x^2}\nonumber \\
\mu_{1,2}\, &=\, \sqrt{\lambda}\, \left(\mu_{1,2}^{(0)} + \sum_{n=2}^{\infty} \mu_{1,2}^{(n)} \lambda^n\right)\label{expansions}
\end{align}
Inserting \eqref{expansions} into \eqref{evsaddles} gives algebraic equations at each order for the $\rho_{1,2}^{(i,n)}$ and $\mu_{1,2}^{(n)}$, which are easily solved. Thus we obtain

\noindent $\boxed{\mathbf{\lambda\ll 1:}}$
\begin{align}
\rho_1 (x)\, &=\, \frac{1}{2\pi g_1^2} \Big(1+2(6\zeta (3))\, g_1^2(g_1^2-g_2^2)\nonumber\\
 &\hspace{0.5in}- 2(20\zeta (5))\, g_1^2(g_1^2-g_2^2)[(3g_1^2+g_2^2)+x^2]+\cdots \Big)\sqrt{\mu_1^2-x^2}\nonumber\\
\mu_1\, &=\, 2 g_1 \Big(1 - 6\zeta (3) g_1^2 (g_1^2 - g_2^2) + 20\zeta (5) g_1^2 (g_1^2 - g_2^2)(4g_1^2 + g_2^2) +\cdots \Big)
\quad . \label{densityexp}
\end{align}
For the second gauge group we have the same expressions but with $g_1\rightarrow g_2$. Note that since $x\sim g_{1,2}$ on the support of the distributions, the two terms in the square brackets in \eqref{densityexp} are of the same order. In appendix \ref{app:expansion} we compute the density up to order $g^{14}$.  At strong coupling, the leading term in $\lambda$ is already implicit in previous work \cite{Rey:2010ry}, which gives for both gauge groups
%
\begin{flalign}\label{strongdist}
&\boxed{\mathbf{\lambda\gg 1:}}\hspace{1in} \rho_{1,2} (x)\, =\, \frac{2}{\pi\mu_{1,2}^2}\sqrt{\mu_{1,2}^2-x^2}&\nonumber\\
&\hspace{1.8in}\mu_1\, =\mu_2\, =\, 2\frac{g_1 g_2}{\sqrt{g_1^2+g_2^2}}\quad .&
\end{flalign}
Note that, in this leading contribution, the two densities are equal, and have the same form as the $\mathcal{N}=4$ density up to a substitution of the coupling. 

We will focus on the totally (anti-)symmetric rank-$k$ representation $\mathcal{R}=S_k/A_k$. The traces over $S_k/A_k$ are given by contour integrals in an auxiliary variable $t$ of two generating functions \cite{Hartnoll:2006is}:
\begin{align}
\label{trace}
\mathrm{Tr}_{S_k}\e^{2\pi a}\, &=\, \frac{1}{2\pi {\rm i}}\oint\dd t\, t^{k-1}\, \prod_{i=1}^N\, \frac{1}{1-t^{-1}\e^{2\pi a_i}}\nonumber\\
 \mathrm{Tr}_{A_k}\e^{2\pi a}\, &=\, \frac{1}{2\pi {\rm i}}\oint\dd t\, \frac{1}{t^{N-k+1}}\, \prod_{i=1}^N\, (t+\e^{2\pi a_i})\quad .
\end{align}
with contours taken anticlockwise around $t=0$. 

We use \eqref{trace} in the matrix integral \eqref{matrixpath}, and take all terms inside the exponential. Then, in the continuum limit, when the integral is dominated by the saddle point densities \eqref{densityexp}\eqref{strongdist} we can write \cite{Hartnoll:2006is}
\begin{align}
W_{S_k/A_k}\, &\equiv \, \frac{1}{\mathrm{dim}({\cal R})} \braket{\mathrm{Tr}_{\cal R}\e^{2\pi a}}_{\text{matrix model}}\nonumber\\
\label{contourint}&= \, \frac{1}{\mathrm{dim}({\cal R})}\oint\dd t\, \exp{N\left( \mp\int\dd x\, \rho_1 (x) \log (1\mp t^{-1}\e^{2\pi x}) + \frac{k}{N} \log t\right)}
\end{align}
where the upper (lower) sign is for the symmetric (antisymmetric) representation. Since we are in the large-$N$ limit, we can evaluate the contour integral \eqref{contourint} using the saddle point method. Differentiating the exponent with respect to $t$, we find the equation determining the saddle points in the $t$-plane:
\begin{align}
\label{saddle}
\int \dd x\,\frac{\rho_1 (x)}{t\,\e^{-2\pi x}\mp 1}\,-\frac{k}{N}\,=\,0\qquad .
\end{align}
The expressions for $\rho_1 (x)$ and $\mu_1$  at weak coupling we know \eqref{densityexp}, so they can be substituted into \eqref{saddle} to yield an expansion in the same way as for \eqref{evsaddles}. We solve this order-by-order for $t$, finding that the saddle point always lies on the real axis. At strong coupling we can use the Wigner distribution \eqref{strongdist}, and again find the saddle on the real axis. Then, evaluating \eqref{contourint} on the saddle, we obtain an expression for the expectation value in each regime. 
\section{The effective coupling}
The full results of our calculation are shown in appendix \ref{app:expansion}. The questions we want to ask are: what is the `effective coupling', in the sense of \eqref{eq:couplingsub}, which we must substitute into the $\mathcal{N}=4$ result \cite{Hartnoll:2006is} to obtain the result for the $\hat{A}_1$ theory? And is this effective coupling the same for all observables in the $SU(2,1|2)$ sector? 

We may indeed construct such an effective coupling. However the answer to the second question is `no': the effective coupling appears to depend on the observable.  In the case of Wilson loops in these higher rank representations, differences at weak coupling only appear at order $\lambda^{10}$ and higher. We quote here the leading terms (we have expanded to order $\lambda^{14}$ in equation \eqref{couplingdiff}). As before the upper (lower) sign is for the symmetric (antisymmetric) loop:

$\boxed{\mathbf{\lambda\ll 1:}}$
\begin{align}\label{anomaly}
&g_{\rm eff}^2 (S_k)-g_{\rm eff}^2 (\Box)\nonumber\\
&=24\, \left(\pm\frac{k}{N}\right)\left(1\pm \frac{k}{N}\right)\, g_1^8\, (g_1^2-g_2^2)\, \left[\, \zeta (2)(20\zeta (5))\,\phantom{\frac{1}{1}}\right.\nonumber\\
&\hspace{1.0in} \left.  -\, 4\zeta (2)^2(20\zeta (5))g_1^2\, -\, \frac{2}{5}\zeta (2)(70\zeta (7))(11g_1^2+5g_2^2)\right] +\mathcal{O}(g^{14})\, .
\end{align}
\eqref{anomaly} is the main result of this note. We see the explicit dependence on the rank $k$ of the representation we consider. The $k$-dependent pre-factor in \eqref{anomaly} is proportional to the quadratic Casimir of the loop representation. More complicated polynomial dependence on $k$ appears at the next order in $g$. Up to the order we have gone to, all of the $k$ depedence can be written as polynomials in the quadratic Casimirs -- this is to be expected since it naturally emerges from products of loop-to-loop propagators. Note the $k\rightarrow N-k$ symmetry in the antisymmetric case, since these two representations are identical. In the $k/N\rightarrow 0$ limit \eqref{anomaly} vanishes and so the effective coupling reduces to that for the fundamental loop \eqref{eq:couplingsub} -- this is because in this limit the leading large-$N$ term is ${\rm Tr}_{S_k/A_k}\e^{2\pi a}=({\rm Tr}_{\Box}\e^{2\pi a})^k+\cdots $. An important property of \eqref{anomaly}, and its full form in the appendix, for our purposes will be that it is entirely proportional to (powers of) $\zeta (2)$. We have grouped the $\zeta$-functions into the combinations coming from the Feynman diagrams \eqref{fandiagrams}.

In the strong coupling limit, as noted in \cite{Rey:2010ry}, the leading form \eqref{strongdist} of the eigenvalue distribution is the same as for the $\mathcal{N}=4$ theory with an effective coupling $1/g^2_{\rm eff}=1/g_1^2+1/g_2^2$. Thus one can immediately see that the coupling substitution in this limit is universal for all expectation values computed through the matrix model \eqref{matrixpath}, so that
\begin{flalign}\label{strongeffcoupling}
&\boxed{\mathbf{\lambda\gg 1:}}\hspace{1.5in}g_{\rm eff}^2 (S_k)-g_{\rm eff}^2 (\Box)\, =\, 0\, +\, \cdots\quad .&
\end{flalign}

Let us stop to discuss the interpretation of our result. We start with weak coupling. The diagrammatic arguments leading to the conjecture of the universal coupling substitution \eqref{eq:couplingsub} are valid for the $\hat{A}_1$ quiver theory formulated on $\mathbb{R}^4$. In the theory on $\mathbb{R}^4$, Feynman diagrams contain both UV and IR divergences, which can both be regulated by (for example) dimensional regularization. However the calculation in this paper is for the theory placed on a round $S^4$. This differs from the flat case in at least two ways. Firstly, there are extra terms which must be added to the Lagrangian in order to preserve supersymmetry on the curved manifold (including a conformal coupling of the adjoint scalars to the scalar curvature of the sphere). Secondly, the compact space introduces, in addition to any other regularization, a hard IR cutoff coming from the smallest eigenvalue of the kinetic operator for each field. 

In order to make a meaningful comparison between our calculations and those coming from Feynman diagrams on flat space, we must take the limit of large radius in our results. However in even in this limit there is no guarantee that this will reproduce the same result as one which treats the theory on $\mathbb{R}^4$ to begin with -- extra contributions could arise from the differences just discussed. That is to say: the regularization procedure and the decompactification limit need not commute. Indeed, for instance, it has been noted in \cite{Russo:2013sba} that placing an $\mathcal{N}=2$ theory on a $S^4$ selects a particular vacuum which remains in the large radius limit. 

In fact \eqref{anomaly} corresponds to exactly such terms. Although our results are explicitly independent of the radius of the sphere, they differ from those obtained from from an \textit{a priori} flat calculation by a discrete contribution beginning at order $\lambda^{14}$. In an upcoming paper \cite{ellinew}, the authors argue that just such contributions (proportional to $\zeta (2)$ and starting at order $\lambda^{14}$) arise when we take into account the IR cutoff introduced by the sphere, which can be viewed as an extra IR regulator for the theory. 

At strong coupling, these effects disappear in the leading order approximation. In particular, this implies that the Wilson loop expectation value is insensitive to whether it is on the sphere or flat space -- it would be interesting to compute the expectation value of the high rank Wilson loop in the orbifold $AdS$ dual with both flat and spherical boundaries, to compare. However note that the dual D-brane solutions are only know for the diagonal combination of Wilson lines in the two gauge groups. The subleading terms in $\lambda$ are much harder to find, although such progress could be made, for example along the lines of \cite{Passerini:2011fe}, and we would like to know if they are non-zero. 

Finally, the discrepancy $g_{\rm eff}^2 (S_k)-g_{\rm eff}^2 (\Box)$ provides us with an interesting interpolating function between strong and weak coupling in the $SU(N)\times SU(N)$ quiver, whose complete form as a function of $\lambda$ it would be interesting and feasible to compute, if a complete large-$N$ solution of the matrix model were found. 
%
%
%
%
%
%
%
%
%
\section{Discussion}
In this paper we have studied one class of observables in the $SU(2,1|2)$ sector of the the $\mathcal{N}=2$ $SU(N)\times SU(N)$ quiver SCFT at both strong and weak coupling: the rank-$k$ (anti-)symmetric equatorial Wilson loop. We confirmed, up to terms that can be explained as IR divergences, that its expectation value can be mapped to the corresponding observable in $\mathcal{N}=4$ SYM by the coupling subsitution conjectured for this sector in \cite{Pomoni:2013poa,Mitev:2014yba}, and found the expansions of the effective coupling in each regime, as functions of the rank $k$ of the representation. We identified an interesting interpolating function which may be calculable at any 't Hooft coupling. 

The $SU(2,1|2)$ sector corresponds (at least at strong coupling) to an $AdS_5\times S^1$ factor in the dual geometry, in the sense that local operators in the sector are dual to string states which classically move within this subspace. Indeed if we take a lesson from $\mathcal{N}=4$ SYM, the dual object for the symmetric loop should be a D-brane with worldvolume entirely on $AdS_5$, while for the antisymmetric loop the situation is less clear. The string dual of the theory is known to be $AdS_5\times S^5/\mathbb{Z}_2$, although the dual to higher rank loops not known for general representations. 

There are still further avenues to explore. Most obviously one can try to generalize our result to $\hat{A}_K$ quiver theories for $K>1$, where a coupling substitution is also conjectured to work, as well as to the rest of the ADE theories. We may also deform these theories by masses for the bifundamental hypermultiplets.

One could look at different regimes in the space of couplings of the $\hat{A}_1$ theory. In particular, a recent paper \cite{Aharony:2015zea} used a specific regime in an attempt to study the 6D (2,0) theory placed on $AdS_5\times S^1$.  In our case this limit corresponds to sending one 't Hooft coupling to infinity while keeping the other fixed, so that a dual description exists in terms of a weakly coupled $SU(2)$ gauge group. It would be interesting to compute Wilson loops there, because it may shed light on the nature of this Seiberg-like duality, as well as the details of the conjectured novel holographic dual of \cite{Aharony:2015zea}. Another limit is to take e.g. $\kappa_1$ to zero while keeping $\lambda$ small, and is a limit described in our calculations. However at generic $\lambda$ the $\kappa_1\rightarrow 0$ limit is disconnected from the those we take in this note, and leads to the strange behaviour of the $N_f=2N$ theory \cite{Fraser:2011qa}. The approach to this limit would be interesting to study. 
\section*{Acknowledgments}
I would like to thank Elli Pomoni for invaluable discussions and for commenting on draft versions, and Daniel Schofield for help with \textit{Mathematica}. My research is implemented under the ``ARISTEIA" action of the ``operational programme education and lifelong learning" and is co-funded by the European Social Fund (ESF) and National Resources. 
\appendix
\section{The full expansions}\label{app:expansion}
In this appendix we quote our full weak coupling results. We have gone up to order $\lambda^{14}$ -- the perturbative expansions were all implemented using simple \textit{Mathematica} functions. First we give the coupling substitution obtained from the localization computation of the Wilson loop in the fundamental representation, up to the order we need - this is an extension to higher order of the result in \citep{Mitev:2014yba}:
\begin{align}
\label{eq:couplingsublong}
&g_{\text{eff}}^2 (\Box) (g_1^2,g_2^2)\, =\, g_1^2-2 g_1^4 \left(g_1^2-g_2^2\right) (6 \zeta (3))+2 g_1^4 \left(g_1^2-g_2^2\right)\left(3 g_1^2+g_2^2\right) (20\zeta (5))\nonumber\\
&+\frac{4}{3} g_1^4 \left(g_1^2-g_2^2\right) \left[3 \left(2 (6\zeta (3))^2+\zeta (2) (20\zeta (5))-4 (70\zeta (7))\right) g_1^4\phantom{\frac{1}{1}}\right.\nonumber\\
&\hspace{1in} \left. -\frac{3}{2} \left(2(6\zeta (3))^2+5 (70\zeta (7))\right) g_1^2 g_2^2+\frac{3}{2} \left(2(6\zeta (3))^2-(70\zeta (7))\right) g_2^4\right]\nonumber\\
&+g_1^4 (g_1^2 - g_2^2) \left[ (252\zeta (9))P^{(12)}_{9} + (70\zeta (7))P^{(12)}_{7} \right.\nonumber\\
&\hspace{2.2in}\left.+ (20\zeta (5))P^{(12)}_{5} + (20\zeta (5))(6\zeta (3))P^{(12)}_{5,3} \right]\nonumber\\
 &+ g_1^4 \left( g_1^2-g_2^2\right) \left[ (924\zeta (11))P^{(14)}_{11}+(252\zeta (9))P^{(14)}_{9} + (70\zeta (7))P^{(14)}_{7}\right.\nonumber\\
 &\hspace{1in} + (70\zeta (7))(6\zeta (3))P^{(14)}_{7,3} + (20\zeta (5))^2P^{(14)}_{5,5}\nonumber\\
&\hspace{1in} \left. + (20\zeta (5))P^{(14)}_{5}+(20\zeta (5))(6\zeta (3))P^{(14)}_{5,3}+(6\zeta (3))^3 P^{(14)}_{3,3,3}\right]
\end{align}
where
\begin{align}
P^{(12)}_{9}&=\frac{2}{3}(65 g_1^6 + 53 g_1^4 g_2^2 + 23 g_1^2 g_2^4 + 3 g_2^6)\nonumber\\
P^{(12)}_{7}&= \frac{8}{5} \zeta (2) g_1^4 (11 g_1^2 + 5 g_2^2)\nonumber\\
P^{(12)}_{5}&= -\frac{64}{5}\zeta(2)^2 g_1^6\nonumber\\
P^{(12)}_{5,3}&= -\frac{4}{5} (15 g_1^6 - 5 g_1^4 g_2^2 + g_1^2 g_2^4 + 5 g_2^6)\nonumber\\
P^{(14)}_{11}&=\frac{2}{15} \left(-915 g_1^8-840 g_1^6 g_2^2-540 g_1^4 g_2^4-165 g_1^2 g_2^6-15 g_2^8\right)\nonumber\\
P^{(14)}_{9}&=\frac{2}{15} \zeta (2)\left(460 g_1^8+340 g_1^6 g_2^2+100 g_1^4 g_2^4\right)\nonumber\\
P^{(14)}_{7}&=\frac{16}{5} \zeta (2)^2 (17g_1^8+g_1^6 g_2^2)\nonumber\\
P^{(14)}_{7,3}&=4\left(48 g_1^8 - 7 g_1^6 g_2^2 - 7 g_1^4 g_2^4 + 11 g_1^2 g_2^6 + 11 g_2^8\right)\nonumber\\
  P^{(14)}_{5,5}&= \frac{2}{15}(855 g_1^8 - 120 g_1^6 g_2^2 - 150 g_1^4 g_2^4 + 210 g_1^2 g_2^6 + 195 g_2^8)\nonumber\\
  P^{(14)}_{5}&= \frac{2}{5} 112 \zeta (2)^3 g_1^8\nonumber\\
  P^{(14)}_{5,3}&= -8\zeta (2) g_1^4 (5 g_1^4 - 4 g_1^2 g_2^2 + g_2^4)\nonumber\\
  P^{(14)}_{3,3,3}&= \frac{2}{15} (-300 g_1^8 + 300 g_1^6 g_2^2 - 300 g_1^4 g_2^4 + 180 g_1^2 g_2^6 - 120 g_2^8)\quad .
\end{align}
In the weak coupling limit, we will be working with the perturbative eigenvalue distribution for the $\hat{A}_1$ quiver, obtained by solving \eqref{evsaddles}, which we quote here
\begin{align}
&\rho_1 (x)\, =\, \frac{1}{2\pi g_1^2} \Bigg(1+2(6\zeta (3))\, g_1^2(g_1^2-g_2^2)\nonumber\\
 &\hspace{1in}\color{brown}{-2(20\zeta (5))\, g_1^2(g_1^2-g_2^2)\Big[(3g_1^2+g_2^2)+x^2\Big]}\nonumber\\
&\color{red}{+\frac{2}{5} g_1^2 (g_1^2-g_2^2) \Big[3 x^4 (70\zeta (7))+2 x^2 (70\zeta (7)) (8 g_1^2+5 g_2^2)}\nonumber\\
&\color{red}{-10 (6\zeta (3))^2 \left(g_1^4+g_2^4\right)+(70\zeta (7)) \left(43 g_1^4+25 g_1^2 g_2^2+5 g_2^4\right)\Big]}\nonumber\\
&\color{blue}{-\frac{2}{21} g_1^2 (g_1^2-g_2^2) \Big[6 x^6 (252\zeta (9))+6 x^4 (252\zeta (9)) (9 g_1^2+7 g_2^2)\nonumber}\\
&\color{blue}{+x^2 \left(-42 (6\zeta (3)) (20\zeta (5)) \left(g_1^4+g_2^4\right)+2 (252\zeta (9)) \left(95 g_1^4+77 g_1^2 g_2^2+35 g_2^4\right)\right)}\nonumber\\
&\color{blue}{ -42 (6\zeta (3)) (20\zeta (5)) P^{(12)}_{x^0,1}+(252\zeta (9)) P^{(12)}_{x^0,2}\Big]}\nonumber\\
&\color{purple}{+\frac{1}{105} g_1^2 (g_1^2-g_2^2) \Big[25 x^8 (924\zeta (11))+50 x^6 (924\zeta (11)) (7 g_1^2+6 g_2^2)}\nonumber\\
&\color{purple}{+6 x^4 \left(-42 (6\zeta (3)) (70\zeta (7)) \left(g_1^4+g_2^4\right)+25 (924\zeta (11)) \left(12 g_1^4+11 g_1^2 g_2^2+7 g_2^4\right)\right)}\nonumber\\
&\color{purple}{+2 x^2 \left(25 (924\zeta (11)) P^{(14)}_{x^2,1}- 210 (20\zeta (5))^2 P^{(14)}_{x^2,2}+84 (6\zeta (3)) (70\zeta (7)) P^{(14)}_{x^2,3}\right)}\nonumber\\
&\color{purple}{+2 \left(5 (924\zeta (11)) P^{(14)}_{x^0,1}+420 (6\zeta (3))^3 P^{(14)}_{x^0,2}\right.}\nonumber\\
&\hspace{1.8in}\color{purple}{\left.-105 (20\zeta (5))^2 P^{(14)}_{x^0,3}-42 (6\zeta (3)) (70\zeta (7)) P^{(14)}_{x^0,4}\right)\Big]}\nonumber\\
&\hspace{4in}+\cdots \Bigg)\sqrt{\mu_1^2-x^2}
\end{align}
where the different colours correspond to different orders of the perturbative expansion, and the $P$'s are the following polynomials in the gauge couplings:
\begin{align}
P^{(12)}_{x^0,1}\, &=\, 9 g_1^6-g_1^6 g_2^2+3 g_1^2 g_2^4+5 g_2^6\nonumber\\
P^{(12)}_{x^0,2}\, &=\, 533 g_1^6+413 g_1^4 g_2^2+161 g_1^2 g_2^4+21 g_2^6\nonumber\\
P^{(14)}_{x^2,1}\, &=\, 109 g_1^6+99 g_1^4 g_2^2+63 g_1^2 g_2^4+21 g_2^6\nonumber\\
P^{(14)}_{x^2,2}\, &=\, 3 g_1^6+g_1^4 g_2^2+g_1^2 g_2^4+3 g_2^6\nonumber\\
P^{(14)}_{x^2,3}\, &=\, 16 g_1^6-3 g_1^4 g_2^2+3 g_1^2 g_2^4+10 g_2^6\nonumber\\
P^{(14)}_{x^0,1}\, &=\, 1636 g_1^8+1461 g_1^6 g_2^2+861 g_1^4 g_2^4+231 g_1^2 g_2^6+21 g_2^8\nonumber\\
P^{(14)}_{x^0,2}\, &=\, 2 g_1^8-g_1^6 g_2^2+2 g_1^4 g_2^4-g_1^2 g_2^6+2 g_2^8\nonumber\\
P^{(14)}_{x^0,3}\, &=\, 41 g_1^8-4 g_1^6 g_2^2+16 g_1^2 g_2^6+13 g_2^8\nonumber\\
P^{(14)}_{x^0,4}\, &=\, 169 g_1^8-11 g_1^6 g_2^2+8 g_1^4 g_2^4+65 g_1^2 g_2^6+55 g_2^8\quad .
\end{align}
The endpoint $\mu_1$ is given by
\begin{align}
&\mu_1\, =\, 2 g_1 \Big(1 - 6\zeta (3) g_1^2 (g_1^2 - g_2^2) + 20\zeta (5) g_1^2 (g_1^2 - g_2^2)(4g_1^2 + g_2^2)\nonumber\\
& +g_1^2 (g_1^2 - g_2^2)\left((70\zeta (7)) \left(-13 g_1^4-7 g_1^2 g_2^2-g_2^4\right)+\frac{1}{2} (6\zeta (3))^2 \left(7 g_1^4-3 g_1^2 g_2^2+4 g_2^4\right)\right)\nonumber\\
&+g_1^2 (g_1^2 - g_2^2)\left((252\zeta (9)) P^{(12)}_9-(6\zeta (3)) (20\zeta (5)) P^{(12)}_{3,5}\right)\nonumber\\
&+g_1^2 (g_1^2 - g_2^2) \left(  (924\zeta (11)) P^{(14)}_{11}+(6\zeta (3))^3 P^{(14)}_{3,3,3}\right.\nonumber\\
&\hspace{2in}\left. + (6\zeta (3)) (70\zeta (7)) P^{(14)}_{3,7}+(20\zeta (5))^2 P^{(14)}_{5,5}  \right)\nonumber\\
& +\cdots \Big)
\end{align}
\begin{align}
P^{(12)}_9&=\left(41 g_1^6+31 g_1^4 g_2^2+11 g_1^2 g_2^4+g_2^6\right)\nonumber\\
P^{(12)}_{3,5}&=\left(34 g_1^6-13 g_1^4 g_2^2+5 g_1^2 g_2^4+10 g_2^6\right)\nonumber\\
P^{(14)}_{11}&=-\frac{1}{2}\left(131 g_1^8+116 g_1^6 g_2^2+66 g_1^4 g_2^4+16 g_1^2 g_2^6+g_2^8\right)\nonumber\\
P^{(14)}_{3,3,3}&=\frac{1}{2}\left(-33 g_1^8+30 g_1^6 g_2^2-33 g_1^4 g_2^4+20 g_1^2 g_2^6-16 g_2^8\right)\nonumber\\
P^{(14)}_{3,7}&=\left(133 g_1^8-32 g_1^6 g_2^2-14 g_1^4 g_2^4+31 g_1^2 g_2^6+22 g_2^8\right)\nonumber\\
P^{(14)}_{5,5}&=\frac{1}{2}\left(158 g_1^8-40 g_1^6 g_2^2-21 g_1^4 g_2^4+41 g_1^2 g_2^6+26 g_2^8\right)\quad .
\end{align}
Note that the operator we are considering explicitly scales exponentially with $N$ in the 't Hooft limit - this can be seen from \eqref{contourint}, where the expression in brackets is $\mathcal{O}(N^0)$. Thus we write $W_{S_k/A_k}=\exp N\, \mathcal{F}_{S_k/A_k} (k/N)$ for some order-one functions $\mathcal{F}_{S_k/A_k}$. For the $\mathcal{N}=4$ theory we obtain
\begin{align}
\label{ne4result}
&\mathcal{F}_{\mathcal{N}=4} (k/N)(g^2)=\nonumber\\
&\pm\hat{C}_2(\mathbf{k}) \left( \frac{(4\pi g)^2}{8} -\frac{(4\pi g)^4}{384}+\frac{(4\pi g)^6}{9216}+\frac{\left(-4+\hat{C}_2(\mathbf{k})\right) (4\pi g)^8}{737280}\right.\nonumber\\
&-\frac{ \left(-13+10 \hat{C}_2(\mathbf{k})\right) (4\pi g)^{10}}{44236800}
 -\frac{\left(495-774 \hat{C}_2(\mathbf{k})+40 \hat{C}_2(\mathbf{k})^2\right) (4\pi g)^{12}}{29727129600}\nonumber\\
&\left. +\frac{\left(3235-8526 \hat{C}_2(\mathbf{k})+1512 \hat{C}_2(\mathbf{k})^2\right) (4\pi g)^{14}}{3329438515200}\right)+\mathcal{O} (g^{16})
\end{align}
where as always the upper (lower) sign is for the symmetric (antisymmetric) loop. $\hat{C}_{2i}$ are the rescaled Casimirs of the representation of the loop:
\begin{align}
\hat{C}_{2i}(\mathbf{k})\, \equiv\, \frac{C_{2i}(\mathbf{k})}{N^{2i}}\, =\, \pm\frac{k}{N}\left( 1\pm\frac{k}{N}\right)^{i-1}
\end{align}
which are order one when $k$ scales with $N$. The main result of this note is the analogous quantity in the $\hat{A}_1$ quiver theory. We have gone up to order $\lambda^{14}$ in the expansion. The result is very long, so it is best to subtract off the result obtained by substituting the coupling substitution \eqref{eq:couplingsublong} into the $\mathcal{N}=4$ result \eqref{ne4result}. The quantity thus obtained is non-zero, indicated that the same coupling subsitution does not work for both the fundamental and higher rank loops.  We find
\begin{align}
&\mathcal{F}_{\hat{A}_1} (k/N)(g_1^2,g_2^2)\, -\, \mathcal{F}_{\mathcal{N}=4} (k/N)(g_{\rm eff}^2 (\Box)(g_1^2,g_2^2)) \nonumber\\
&=\zeta (2)^2\, g_1^8\, (g_1^2-g_2^2)\, \hat{C}_2 (\mathbf{k})^2\, \left(\phantom{\frac{1}{1}} 288(20\zeta (5))\, \right.\nonumber\\
&\hspace{1in}\left. -\, 2304\zeta (2)(20\zeta (5))g_1^2\, -\, \frac{576}{5}(70\zeta (7))(11g_1^2+5g_2^2)\right.\nonumber\\
&\hspace{1in}-\, \frac{1152}{5}\left[ -74\zeta (2)^2(20\zeta (5)) g_1^4
 +\frac{5}{2}(5g_1^4-4g_1^2g_2^2+g_2^4)(6\zeta(3))(20\zeta (5))\right.\nonumber\\
&\hspace{1.7in} +\zeta (2) (70\zeta (7))\, g_1^2 \left((-41+12\hat{C}_2 (\mathbf{k}))g_1^2-20g_2^2\right)\nonumber\\ 
&\hspace{1.7in}\left.\left. -\frac{5}{6} (23g_1^4+17g_1^2g_2^2+5g_2^4)(252\zeta (9))   \right] \phantom{\frac{1}{1}}\right)+\mathcal{O} (g^{16})\; .
\end{align}
Given this discrepancy, we can compute the correct renormalized effective coupling for the higher rank loops. Again taking the difference with the result for the fundamental, we have that
\begin{align}
\label{couplingdiff}
&g_{\rm eff}^2 (S_k)-g_{\rm eff}^2 (\Box)\nonumber\\
&=24\, g_1^8\, (g_1^2-g_2^2)\, \hat{C}_2(\mathbf{k})\, \left[\phantom{\frac{1}{1}}  \zeta (2)(20\zeta (5))\, \right.\nonumber\\
&\hspace{1in}-\, 4\zeta (2)^2(20\zeta (5))g_1^2\, -\, \frac{2}{5}\zeta (2)(70\zeta (7))(11g_1^2+5g_2^2)\nonumber\\
&\hspace{1in}+\frac{96}{5}\zeta (2)^3(20\zeta (5)) g_1^4
 +2\zeta (2)(6\zeta(3))(20\zeta (5))(5g_1^4-4g_1^2g_2^2+g_2^4)\nonumber\\
&\hspace{1.6in} +\frac{4}{5}\zeta (2)^2 (70\zeta (7))\, g_1^2 \left( (19-12\hat{C}_2(\mathbf{k}))g_1^2+10 g_2^2\right)\nonumber\\ 
&\hspace{1.6in}\left. +\frac{2}{3} \zeta (2)(252\zeta (9)) (23g_1^4+17g_1^2g_2^2+5g_2^4)  \right]+\mathcal{O} (g^{16})\; .
\end{align}
\bibliographystyle{utphys}
\bibliography{a1ref.bib}

\providecommand{\href}[2]{#2}\begingroup\raggedright\begin{thebibliography}{10}

\bibitem{Gaiotto:2009we}
D.~Gaiotto, ``{N=2 dualities},''
  \href{http://dx.doi.org/10.1007/JHEP08(2012)034}{{\em JHEP} {\bfseries 1208}
  (2012) 034},
\href{http://arxiv.org/abs/0904.2715}{{\ttfamily arXiv:0904.2715 [hep-th]}}.

\bibitem{Tachikawa:2013kta}
Y.~Tachikawa, ``{N=2 supersymmetric dynamics for pedestrians},''
  \href{http://dx.doi.org/10.1007/978-3-319-08822-8}{{\em Lect.Notes Phys.}
  {\bfseries 890} (2013) 2014},
\href{http://arxiv.org/abs/1312.2684}{{\ttfamily arXiv:1312.2684 [hep-th]}}.

\bibitem{Grana:2001xn}
M.~Grana and J.~Polchinski, ``{Gauge / gravity duals with holomorphic
  dilaton},'' \href{http://dx.doi.org/10.1103/PhysRevD.65.126005}{{\em
  Phys.Rev.} {\bfseries D65} (2002) 126005},
\href{http://arxiv.org/abs/hep-th/0106014}{{\ttfamily arXiv:hep-th/0106014
  [hep-th]}}.

\bibitem{Lin:2004nb}
H.~Lin, O.~Lunin, and J.~M. Maldacena, ``{Bubbling AdS space and 1/2 BPS
  geometries},'' \href{http://dx.doi.org/10.1088/1126-6708/2004/10/025}{{\em
  JHEP} {\bfseries 0410} (2004) 025},
\href{http://arxiv.org/abs/hep-th/0409174}{{\ttfamily arXiv:hep-th/0409174
  [hep-th]}}.

\bibitem{Gaiotto:2009gz}
D.~Gaiotto and J.~Maldacena, ``{The Gravity duals of N=2 superconformal field
  theories},'' \href{http://dx.doi.org/10.1007/JHEP10(2012)189}{{\em JHEP}
  {\bfseries 1210} (2012) 189},
\href{http://arxiv.org/abs/0904.4466}{{\ttfamily arXiv:0904.4466 [hep-th]}}.

\bibitem{Gadde:2009dj}
A.~Gadde, E.~Pomoni, and L.~Rastelli, ``{The Veneziano Limit of N = 2
  Superconformal QCD: Towards the String Dual of N = 2 SU(N(c)) SYM with N(f) =
  2 N(c)},''
\href{http://arxiv.org/abs/0912.4918}{{\ttfamily arXiv:0912.4918 [hep-th]}}.

\bibitem{ReidEdwards:2010qs}
R.~Reid-Edwards and B.~Stefanski, ``{On Type IIA geometries dual to N = 2
  SCFTs},'' \href{http://dx.doi.org/10.1016/j.nuclphysb.2011.04.002}{{\em
  Nucl.Phys.} {\bfseries B849} (2011) 549--572},
\href{http://arxiv.org/abs/1011.0216}{{\ttfamily arXiv:1011.0216 [hep-th]}}.

\bibitem{Colgain:2011hb}
E.~O~Colgain and B.~Stefa\'{n}ski, ``{A search for AdS5 X S2 IIB supergravity
  solutions dual to N = 2 SCFTs},''
  \href{http://dx.doi.org/10.1007/JHEP10(2011)061}{{\em JHEP} {\bfseries 1110}
  (2011) 061},
\href{http://arxiv.org/abs/1107.5763}{{\ttfamily arXiv:1107.5763 [hep-th]}}.

\bibitem{Aharony:2012tz}
O.~Aharony, L.~Berdichevsky, and M.~Berkooz, ``{4d N=2 superconformal linear
  quivers with type IIA duals},''
  \href{http://dx.doi.org/10.1007/JHEP08(2012)131}{{\em JHEP} {\bfseries 1208}
  (2012) 131},
\href{http://arxiv.org/abs/1206.5916}{{\ttfamily arXiv:1206.5916 [hep-th]}}.

\bibitem{Stefanski:2013osa}
B.~Stefa\'{n}ski, ``{Supermembrane actions for Gaiotto-Maldacena
  backgrounds},'' \href{http://dx.doi.org/10.1016/j.nuclphysb.2014.03.028}{{\em
  Nucl.Phys.} {\bfseries B883} (2014) 581--597},
\href{http://arxiv.org/abs/1308.2789}{{\ttfamily arXiv:1308.2789 [hep-th]}}.

\bibitem{Kachru:1998ys}
S.~Kachru and E.~Silverstein, ``{4-D conformal theories and strings on
  orbifolds},'' \href{http://dx.doi.org/10.1103/PhysRevLett.80.4855}{{\em
  Phys.Rev.Lett.} {\bfseries 80} (1998) 4855--4858},
\href{http://arxiv.org/abs/hep-th/9802183}{{\ttfamily arXiv:hep-th/9802183
  [hep-th]}}.

\bibitem{Lawrence:1998ja}
A.~E. Lawrence, N.~Nekrasov, and C.~Vafa, ``{On conformal field theories in
  four-dimensions},''
  \href{http://dx.doi.org/10.1016/S0550-3213(98)00495-7}{{\em Nucl.Phys.}
  {\bfseries B533} (1998) 199--209},
\href{http://arxiv.org/abs/hep-th/9803015}{{\ttfamily arXiv:hep-th/9803015
  [hep-th]}}.

\bibitem{Gadde:2010zi}
A.~Gadde, E.~Pomoni, and L.~Rastelli, ``{Spin Chains in N=2 Superconformal
  Theories: From the $Z_2$ Quiver to Superconformal QCD},''
  \href{http://dx.doi.org/10.1007/JHEP06(2012)107}{{\em JHEP} {\bfseries 1206}
  (2012) 107},
\href{http://arxiv.org/abs/1006.0015}{{\ttfamily arXiv:1006.0015 [hep-th]}}.

\bibitem{Pomoni:2011jj}
E.~Pomoni and C.~Sieg, ``{From N=4 gauge theory to N=2 conformal QCD:
  three-loop mixing of scalar composite operators},''
\href{http://arxiv.org/abs/1105.3487}{{\ttfamily arXiv:1105.3487 [hep-th]}}.

\bibitem{Liendo:2011xb}
P.~Liendo, E.~Pomoni, and L.~Rastelli, ``{The Complete One-Loop Dilation
  Operator of N=2 SuperConformal QCD},''
  \href{http://dx.doi.org/10.1007/JHEP07(2012)003}{{\em JHEP} {\bfseries 1207}
  (2012) 003},
\href{http://arxiv.org/abs/1105.3972}{{\ttfamily arXiv:1105.3972 [hep-th]}}.

\bibitem{Pomoni:2013poa}
E.~Pomoni, ``{Integrability in N=2 superconformal gauge theories},''
  \href{http://dx.doi.org/10.1016/j.nuclphysb.2015.01.006}{{\em Nucl.Phys.}
  {\bfseries B893} (2015) 21--53},
\href{http://arxiv.org/abs/1310.5709}{{\ttfamily arXiv:1310.5709 [hep-th]}}.

\bibitem{Pestun:2007rz}
V.~Pestun, ``{Localization of gauge theory on a four-sphere and supersymmetric
  Wilson loops},'' \href{http://dx.doi.org/10.1007/s00220-012-1485-0}{{\em
  Commun.Math.Phys.} {\bfseries 313} (2012) 71--129},
\href{http://arxiv.org/abs/0712.2824}{{\ttfamily arXiv:0712.2824 [hep-th]}}.

\bibitem{Russo:2012ay}
J.~Russo and K.~Zarembo, ``{Large N Limit of N=2 SU(N) Gauge Theories from
  Localization},'' \href{http://dx.doi.org/10.1007/JHEP10(2012)082}{{\em JHEP}
  {\bfseries 1210} (2012) 082},
\href{http://arxiv.org/abs/1207.3806}{{\ttfamily arXiv:1207.3806 [hep-th]}}.

\bibitem{Buchel:2013id}
A.~Buchel, J.~G. Russo, and K.~Zarembo, ``{Rigorous Test of Non-conformal
  Holography: Wilson Loops in N=2* Theory},''
  \href{http://dx.doi.org/10.1007/JHEP03(2013)062}{{\em JHEP} {\bfseries 1303}
  (2013) 062},
\href{http://arxiv.org/abs/1301.1597}{{\ttfamily arXiv:1301.1597 [hep-th]}}.

\bibitem{Russo:2013qaa}
J.~G. Russo and K.~Zarembo, ``{Evidence for Large-N Phase Transitions in N=2*
  Theory},'' \href{http://dx.doi.org/10.1007/JHEP04(2013)065}{{\em JHEP}
  {\bfseries 1304} (2013) 065},
\href{http://arxiv.org/abs/1302.6968}{{\ttfamily arXiv:1302.6968 [hep-th]}}.

\bibitem{Russo:2013kea}
J.~Russo and K.~Zarembo, ``{Massive N=2 Gauge Theories at Large N},''
  \href{http://dx.doi.org/10.1007/JHEP11(2013)130}{{\em JHEP} {\bfseries 1311}
  (2013) 130},
\href{http://arxiv.org/abs/1309.1004}{{\ttfamily arXiv:1309.1004 [hep-th]}}.

\bibitem{Russo:2013sba}
J.~Russo and K.~Zarembo, ``{Localization at Large N},''
\href{http://arxiv.org/abs/1312.1214}{{\ttfamily arXiv:1312.1214 [hep-th]}}.

\bibitem{Mitev:2014yba}
V.~Mitev and E.~Pomoni, ``{The Exact Effective Couplings of 4D N=2 gauge
  theories},''
\href{http://arxiv.org/abs/1406.3629}{{\ttfamily arXiv:1406.3629 [hep-th]}}.

\bibitem{Drukker:2005kx}
N.~Drukker and B.~Fiol, ``{All-genus calculation of Wilson loops using
  D-branes},'' \href{http://dx.doi.org/10.1088/1126-6708/2005/02/010}{{\em
  JHEP} {\bfseries 0502} (2005) 010},
\href{http://arxiv.org/abs/hep-th/0501109}{{\ttfamily arXiv:hep-th/0501109
  [hep-th]}}.

\bibitem{Yamaguchi:2006te}
S.~Yamaguchi, ``{Bubbling geometries for half BPS Wilson lines},''
  \href{http://dx.doi.org/10.1142/S0217751X07035070}{{\em Int.J.Mod.Phys.}
  {\bfseries A22} (2007) 1353--1374},
\href{http://arxiv.org/abs/hep-th/0601089}{{\ttfamily arXiv:hep-th/0601089
  [hep-th]}}.

\bibitem{Hartnoll:2006hr}
S.~A. Hartnoll and S.~P. Kumar, ``{Multiply wound Polyakov loops at strong
  coupling},'' \href{http://dx.doi.org/10.1103/PhysRevD.74.026001}{{\em
  Phys.Rev.} {\bfseries D74} (2006) 026001},
\href{http://arxiv.org/abs/hep-th/0603190}{{\ttfamily arXiv:hep-th/0603190
  [hep-th]}}.

\bibitem{Yamaguchi:2006tq}
S.~Yamaguchi, ``{Wilson loops of anti-symmetric representation and
  D5-branes},'' \href{http://dx.doi.org/10.1088/1126-6708/2006/05/037}{{\em
  JHEP} {\bfseries 0605} (2006) 037},
\href{http://arxiv.org/abs/hep-th/0603208}{{\ttfamily arXiv:hep-th/0603208
  [hep-th]}}.

\bibitem{Gomis:2006sb}
J.~Gomis and F.~Passerini, ``{Holographic Wilson Loops},''
  \href{http://dx.doi.org/10.1088/1126-6708/2006/08/074}{{\em JHEP} {\bfseries
  0608} (2006) 074},
\href{http://arxiv.org/abs/hep-th/0604007}{{\ttfamily arXiv:hep-th/0604007
  [hep-th]}}.

\bibitem{Okuyama:2006jc}
K.~Okuyama and G.~W. Semenoff, ``{Wilson loops in N=4 SYM and fermion
  droplets},'' \href{http://dx.doi.org/10.1088/1126-6708/2006/06/057}{{\em
  JHEP} {\bfseries 0606} (2006) 057},
\href{http://arxiv.org/abs/hep-th/0604209}{{\ttfamily arXiv:hep-th/0604209
  [hep-th]}}.

\bibitem{Hartnoll:2006is}
S.~A. Hartnoll and S.~P. Kumar, ``{Higher rank Wilson loops from a matrix
  model},'' \href{http://dx.doi.org/10.1088/1126-6708/2006/08/026}{{\em JHEP}
  {\bfseries 0608} (2006) 026},
\href{http://arxiv.org/abs/hep-th/0605027}{{\ttfamily arXiv:hep-th/0605027
  [hep-th]}}.

\bibitem{Gomis:2006im}
J.~Gomis and F.~Passerini, ``{Wilson Loops as D3-Branes},''
  \href{http://dx.doi.org/10.1088/1126-6708/2007/01/097}{{\em JHEP} {\bfseries
  0701} (2007) 097},
\href{http://arxiv.org/abs/hep-th/0612022}{{\ttfamily arXiv:hep-th/0612022
  [hep-th]}}.

\bibitem{Faraggi:2014tna}
A.~Faraggi, J.~T. Liu, L.~A. Pando~Zayas, and G.~Zhang, ``{One-loop structure
  of higher rank Wilson loops in AdS/CFT},''
  \href{http://dx.doi.org/10.1016/j.physletb.2014.11.060}{{\em Phys.Lett.}
  {\bfseries B740} (2015) 218--221},
\href{http://arxiv.org/abs/1409.3187}{{\ttfamily arXiv:1409.3187 [hep-th]}}.

\bibitem{Fraser:2011qa}
B.~Fraser and S.~P. Kumar, ``{Large rank Wilson loops in N=2 superconformal QCD
  at strong coupling},'' \href{http://dx.doi.org/10.1007/JHEP03(2012)077}{{\em
  JHEP} {\bfseries 1203} (2012) 077},
\href{http://arxiv.org/abs/1112.5182}{{\ttfamily arXiv:1112.5182 [hep-th]}}.

\bibitem{Chen-Lin:2015dfa}
X.~Chen-Lin and K.~Zarembo, ``{Higher Rank Wilson Loops in N = 2*
  Super-Yang-Mills Theory},''
\href{http://arxiv.org/abs/1502.01942}{{\ttfamily arXiv:1502.01942 [hep-th]}}.

\bibitem{Passerini:2011fe}
F.~Passerini and K.~Zarembo, ``{Wilson Loops in N=2 Super-Yang-Mills from
  Matrix Model},'' \href{http://dx.doi.org/10.1007/JHEP10(2011)065,
  10.1007/JHEP09(2011)102}{{\em JHEP} {\bfseries 1109} (2011) 102},
\href{http://arxiv.org/abs/1106.5763}{{\ttfamily arXiv:1106.5763 [hep-th]}}.

\bibitem{Broadhurst:1985vq}
D.~J. Broadhurst, ``{Evaluation of a Class of Feynman Diagrams for All Numbers
  of Loops and Dimensions},''
\href{http://dx.doi.org/10.1016/0370-2693(85)90340-5}{{\em Phys.Lett.}
  {\bfseries B164} (1985) 356}.

\bibitem{Rey:2010ry}
S.-J. Rey and T.~Suyama, ``{Exact Results and Holography of Wilson Loops in N=2
  Superconformal (Quiver) Gauge Theories},''
  \href{http://dx.doi.org/10.1007/JHEP01(2011)136}{{\em JHEP} {\bfseries 1101}
  (2011) 136},
\href{http://arxiv.org/abs/1001.0016}{{\ttfamily arXiv:1001.0016 [hep-th]}}.

\bibitem{ellinew}
V.~Mitev and E.~Pomoni.
\newblock To appear.

\bibitem{Aharony:2015zea}
O.~Aharony, M.~Berkooz, and S.-J. Rey, ``{Rigid Holography and Six-Dimensional
  N=(2,0) Theories on $AdS_5 \times S^1$},''
\href{http://arxiv.org/abs/1501.02904}{{\ttfamily arXiv:1501.02904 [hep-th]}}.

\end{thebibliography}\endgroup
\end{document}